\documentclass[journal]{IEEEtran}
\usepackage{algpseudocode}
\usepackage{algorithm}
\usepackage{cite}
\ifCLASSINFOpdf
   \usepackage[pdftex]{graphicx}
\else
   \usepackage[dvips]{graphicx}
\fi
\usepackage{color}
\usepackage{xcolor}

\ifCLASSOPTIONcompsoc
 \usepackage[caption=false,font=normalsize,labelfont=sf,textfont=sf]{subfig}
\else
 \usepackage[caption=false,font=footnotesize]{subfig}
\fi

\usepackage{bm}
\usepackage{bbm}
\usepackage{amsmath}
\usepackage{amsfonts}
\usepackage{cuted}

\usepackage{tabularx}
\usepackage{threeparttable} 
\usepackage{stfloats}
\usepackage{enumitem}
\usepackage[normalem]{ulem}

\usepackage{hyperref}
\hypersetup{
    colorlinks=true,
    linkcolor=black,
    filecolor=magenta,      
    urlcolor=teal,
    citecolor=black,
    pdftitle={GD_compensation},
    }
\usepackage{xr}
\makeatletter

\newcommand{\tr}[1]{{\mathrm{tr}\left({#1} \right)}}

\newcommand{\E}[1]{{\mathbb{E}\left\{{#1} \right\}}}

\newcommand{\PC}{{\bm{\mathcal{P}}}}
\newcommand{\SC}{{\bm{\mathcal{S}}}}

\newcommand{\tautau}{{\bm{{\tau}}}}
\newcommand{\bbeta}{{\bm{{\beta}}}}
\newcommand{\barbeta}{{\bar{\bm{{\beta}}}}}
\newcommand{\barCD}{{\overline{\bm{{\mathrm{CD}}}}}}

\newcommand{\D}{{\mathbf{D}}}
\newcommand{\Dinv}{{\mathbf{D}^{-1}}}

\newcommand{\Rp}{{\mathbf{R}_P}}
\newcommand{\Rs}{{\mathbf{R}_S}}
\newcommand{\LP}{\mathrm{LP}}
\newcommand{\ttot}{\bm{\tau}_\textrm{tot}}
\newcommand\lenS{L_S}
\newcommand{\nclad}{n_\textrm{clad}}
\newcommand{\sigDeln}{\sigma_{\Delta n}}
\mathchardef\mhyphen="2D 
\newcommand{\linter}{{L_\textrm{inter}}}

\newcommand\norm[1]{\left\lVert#1\right\rVert}

\DeclareMathOperator*{\minimize}{minimize}
\DeclareMathOperator*{\subjectto}{subject~to}

\begin{document}
\title{
Closed-Form Statistics and Design of Mode-Division-Multiplexing Systems Employing Group-Delay Compensation and Mode Permutation
}

\author{Anirudh~Vijay,~\IEEEmembership{Student~Member,~IEEE,} 
Nika~Zahedi,
Oleksiy~Krutko,~\IEEEmembership{Student~Member,~IEEE,}\\Rebecca~Refaee and Joseph~M.~Kahn,~\IEEEmembership{Fellow,~IEEE}%
\thanks{
The authors are with the E.L Ginzton Laboratory, Department of Electrical Engineering, Stanford University, Stanford, CA 94305 (email: avijay@stanford.edu; nzahedi@stanford.edu; oleksiyk@stanford.edu; becca24@stanford.edu; jmk@ee.stanford.edu).

© 2025 IEEE. Personal use of this material is permitted. Permission from IEEE must be obtained for all other uses, in any current or future media, including reprinting/republishing this material for advertising or promotional purposes, creating new collective works, for resale or redistribution to servers or lists, or reuse of any copyrighted component of this work in other works.}%
}

\markboth{Journal of Lightwave Technology}%
{Author \MakeLowercase{\textit{et al.}}: Insert Paper Title here}

\maketitle

\begin{abstract}
Excessive accumulation of group-delay (GD) spread increases computational complexity and affects tracking of receiver-based multi-input multi-output signal processing, posing challenges to long-haul mode-division multiplexing in multimode fiber (MMF). 
GD compensation, which involves periodically exchanging propagating signals between modes with lower and higher GDs, can potentially reduce GD spread. 
In this work, we investigate two GD compensation schemes: conventional compensation, which alternates fiber types with opposite GD orderings, and self-compensation, which employs a single fiber type with periodically inserted mode permuters. 
We provide analytical expressions for GD statistics in compensated systems with arbitrary MMF types and mode permuters, accounting for random inter-group coupling, mode scrambling, and refractive index errors. 
To enhance the effectiveness of GD compensation over the C-band, we propose optimized graded-index depressed-cladding MMFs with index profiles tailored to control mode-dependent chromatic dispersion. 
We also analyze the impact of fiber fabrication errors and explore mitigation strategies based on GD characterization and sorting. 
Design examples and numerical simulations demonstrate improved system performance, highlighting the trade-offs between compensation effectiveness, system complexity, and transmission losses.
For a 5000 km MMF link, the design examples achieve a GD standard deviation (STD) of 438$~\textrm{ps}$ (effectively 6.2$~\textrm{ps}/\sqrt{\textrm{km}}$) and 257$~\textrm{ps}$ (effectively 3.6$~\textrm{ps}/\sqrt{\textrm{km}}$) for conventional and self-compensation schemes, respectively, under ideal conditions. 
In the presence of random inter-group coupling (characteristic coupling length = 500 km), the GD STD increases to 803$~\textrm{ps}$ (11.4$~\textrm{ps}/\sqrt{\textrm{km}}$) and 575$~\textrm{ps}$ (8.1$~\textrm{ps}/\sqrt{\textrm{km}}$), respectively. 
Additionally, with refractive-index errors (error STD = $10^{-5}$), the GD STD further increases to 1.28$~\textrm{ns}$ (18.1$~\textrm{ps}/\sqrt{\textrm{km}}$) for both schemes.

\end{abstract}

\IEEEpeerreviewmaketitle

\section{Introduction}

\IEEEPARstart{S}{pace-division} 
multiplexing (SDM) enhances capacity, integration, and power efficiency in long-haul optical coherent communication systems \cite{richardson_space-division_2013,essiambre_capacity_2012,puttnam_space-division_2021}. 
It can be implemented using parallel single-mode fibers (SMFs), uncoupled-core or coupled-core multi-core fibers (MCFs), or multi-mode fibers (MMFs). 
Among these options, mode-division multiplexing (MDM) in MMFs is particularly attractive for achieving the highest level of integration \cite{winzer_chapter_2013}, enabling amplification with fewer pump modes than signal modes \cite{srinivas_efficient_2023}.

Graded-index (GI) MMFs are preferred for long-haul transmission when the number of spatial and polarization modes exceeds six ($D > 6$) due to their inherently low group-delay (GD) standard deviation (STD) \cite{jensen_demonstration_2015}, a key measure of modal dispersion. 
Reducing the end-to-end GD STD is desirable as it decreases the digital signal processing (DSP) complexity at the receiver.

GD management strategies in the literature can be broadly subdivided into two classes.
The first class focuses on the design of fibers with favorable intrinsic GD properties. 
This includes engineering the refractive index profile to achieve low uncoupled GDs while promoting mode coupling and optimizing wavelength-dependent characteristics such as chromatic dispersion (CD) \cite{sillard_low-differential-mode-group-delay_2016}.
The second class involves system-level techniques to manage GD accumulation, such as mode scrambling to enhance mode coupling \cite{arik_delay_2015, krutko_ultra-low-loss_2024} or GD compensation by concatenating fibers with complementary GD orderings. 
Recently, mode permutation to exchange power between slow and fast mode groups has been shown to reduce end-to-end GD STD, both theoretically \cite{vijay_modal_2025} and experimentally \cite{shibahara_long-haul_2020, sciullo_enhancing_2024, gao_novel_2024}.

This paper focuses on GD compensation, studying two approaches: concatenating multiple fiber types (referred to as \textit{conventional compensation}) and using a single fiber type with mode permutation (referred to as \textit{self-compensation}).
Achieving a sufficiently low end-to-end GD STD requires the design of special fibers having modal GDs suitable for these compensation schemes. The systematic design of fibers for compensation involves optimizing the refractive index profile to achieve specific uncoupled GD values, as proposed in prior research \cite{maruyama_two_2014, gruner-nielsen_few_2012, ferreira_design_2013, mori_few-mode_2014}.

Several challenges impact the effectiveness of GD compensation, necessitating careful consideration in system design.
One key challenge is the wavelength dependence of relevant fiber GD properties, especially mode-dependent chromatic dispersion (MDCD). 
A large MDCD can result in significant variations in effective GD STD across different wavelength channels.
This is undesirable because receiver DSP complexity is dictated by the channel with the highest GD STD. 
Fiber designs should therefore aim to minimize MDCD, thus we propose incorporating pedestal-like features into the refractive index profile, similar to techniques proposed in the literature \cite{gerome_theoretical_2006,riishede_inverse_2008}.
Fiber fabrication errors also pose a challenge to GD compensation \cite{gruner-nielsen_few_2012}. 
It is essential to assess their impact and design fibers with GD properties that maintain effective compensation despite imperfect fabrication.
During system integration, manufactured fibers can be characterized, and fibers with compatible GD orderings can be selected for use in the same span to enhance compensation efficacy.

An important system-level challenge arises from random distributed inter-group coupling in MMFs. 
Since GD compensation relies on controlled and complete power exchanges between fiber segments, random coupling can undermine compensation efficacy \cite{arik_delay_2015}. 
To address this, it is crucial to quantify the impact of inter-group coupling on the effective GD STD and develop fiber and system designs that are resilient to this effect.
One proposed mitigation strategy is to shorten the fiber segments used in GD compensation to reduce the influence of random inter-group coupling.

In this paper, we present a generalized approach to studying both
conventional compensation and self-compensation of GD. 
We derive analytical expressions for end-to-end GD STD, incorporating the effects of mode scrambling and random inter-group coupling. 
For both compensation schemes, we provide design examples of optimized GI-MMFs supporting $12$ spatial and polarization modes and evaluate their GD statistics through extensive numerical simulations. 
We propose parameterized fiber designs and carefully select a cost function that accounts for the system-level effects of modal GDs and MDCDs. 
Using numerical optimization techniques such as particle swarm optimization, we systematically search for fiber designs that minimize the effective GD STD after compensation while ensuring low MDCD and robustness to random inter-group coupling and fabrication errors. 
Additionally, we analyze the impact of fabrication errors and introduce a semi-analytic penalty term in the GD STD expression. Our findings indicate that achieving GD STD values competitive with coupled-core MCFs requires stringent fabrication tolerances.

The rest of the paper is organized as follows: 
Section \ref{sec:Analytical} presents the modeling of GD compensation systems, analytical expressions for GD STD in such systems in the presence of random inter-group coupling, and numerical simulations to validate the analytical expressions.
Section \ref{sec:FiberDesign} presents a fiber design strategy for MMF links supporting six spatial modes in the C-band, design examples for both conventional and self-compensation schemes, and numerical simulations to evaluate the end-to-end GD STD.
Sections \ref{sec:Discussion} and \ref{sec:Conclusion} present discussion and conclusions, respectively.

\begin{figure*}[ht]
\subfloat{\label{fig:1a}}
\subfloat{\label{fig:1b}}
\subfloat{\label{fig:1c}}
\centering
\includegraphics[width=0.8\textwidth]{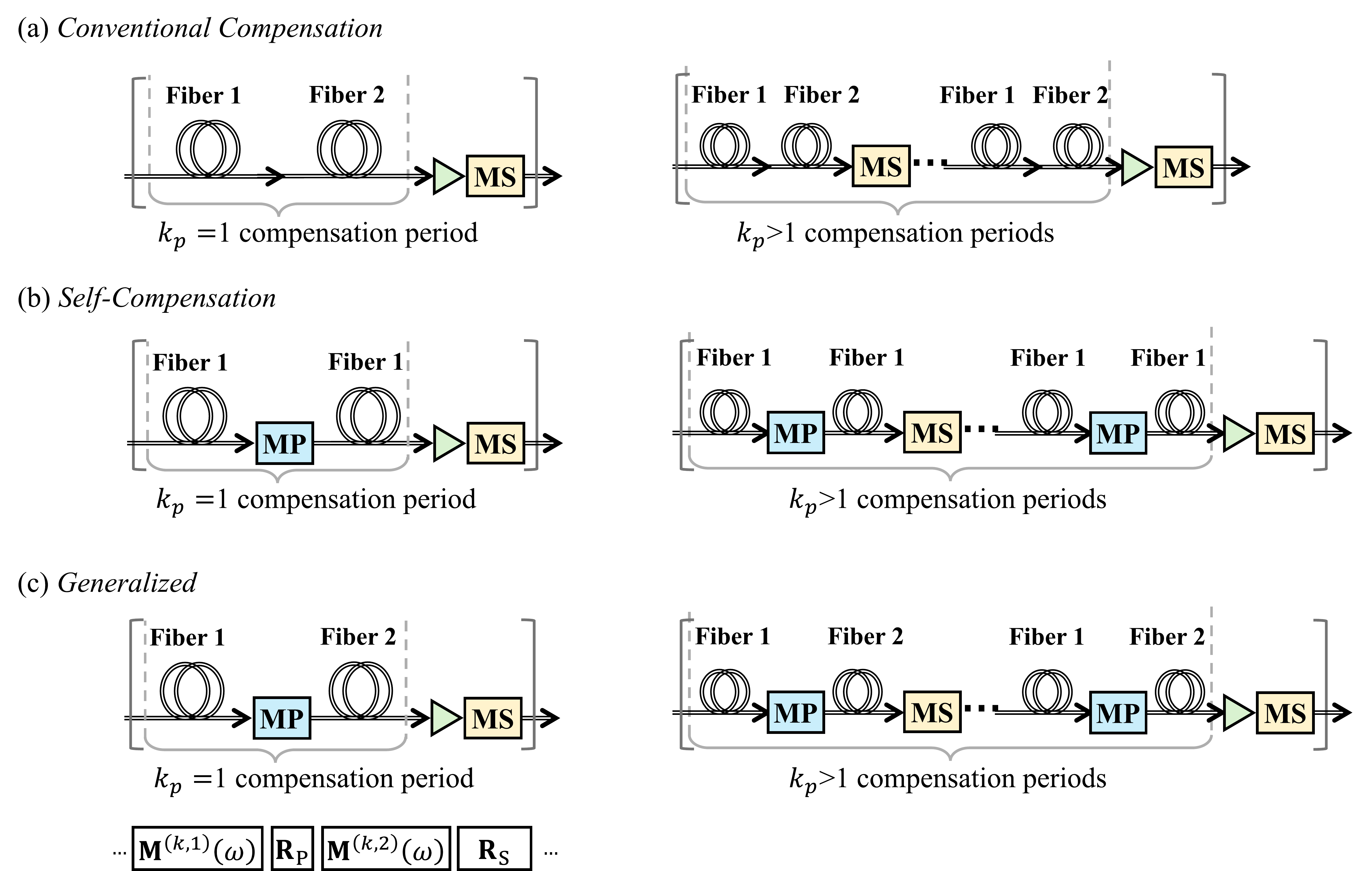}
\caption{Diagrams of one span of a long-haul MDM transmission link with periodic amplification and GD compensation.
Each fiber span can have one or more compensation periods.
(a) Conventional compensation scheme employing two fiber types and mode scramblers; 
(b) Self-compensation scheme employing one fiber type, mode permuters and mode scramblers;
(c) Generalized compensation scheme employing up to two fiber types, mode permuters and mode scramblers. 
MP: Mode Permuter, MS: Mode Scrambler.}
\label{fig:sys_arch}
\end{figure*}

\section{Group-Delay Compensation in MMF Systems}
\label{sec:Analytical}

GD compensation schemes for long-haul MDM transmission systems employing MMFs with periodic amplification and mode scrambling are shown in Fig.~\ref{fig:sys_arch}.
We define a \textit{span} as a portion of the link between successive amplifier subsystems, each including both an inline amplifier and a mode scrambler. 
Each span is of length $\lenS$ and has identical statistical properties. {We define a \textit{segment} as a piece of fiber of a given type.
We define a \textit{compensation period} as a pair of fiber segments whose group delays ideally cancel.\footnote{In Section \ref{subsec:extension}, we 
briefly consider a compensation scheme employing three fiber segments per compensation period.}
A span may contain one or more compensation periods, with the number of periods per span denoted by $k_p$.

The conventional compensation scheme (Fig.~\ref{fig:1a}) uses two fiber types per span whose modal GDs, relative to the mode-averaged GD, are ideally equal and opposite. 
The self-compensation scheme  (Fig.~\ref{fig:1b}) uses one fiber type and a mode permuter, and ideally, the permuter exchanges power between the fastest and the slowest mode groups.
We study both schemes within a generalized framework (Fig.~\ref{fig:1c}) that uses up to two fiber types and an optional mode permuter.
In this section, we provide analytical expressions for the GD STD in systems employing GD compensation and present simulation results that validate these expressions.

As shown in Fig.~\ref{fig:1c}, in the generalized Jones representation for $D$ spatial and polarization modes, the matrices $\mathbf{M}_{k,1}(\omega)$, and $\mathbf{M}_{k,2}(\omega)$ denote the $D\times D$ frequency-dependent complex-baseband random electric field transfer matrices for the two MMF segments in the $k$th span (we are assuming $k_p = 1$ compensation periods per span for the time being). $\Rp$ and $\Rs$ denote the $D\times D$ frequency-independent deterministic transfer matrix of the mode permuters and scramblers, respectively. 
The $D$ modes are grouped into $N_g$ mode groups; modes in a mode group have nearly equal propagation constants. The set of modes in the $i$th mode group is denoted by $\mathcal{M}_i$ and the degeneracy is given by $d_i=\left | \mathcal{M}_i \right|$, where $\left | \cdot \right|$ represents the cardinality. 

\subsection{Analysis of GD Spread in GD Compensation Systems}
Following the analysis in \cite{vijay_modal_2025}, we define mode-group-averaged $N_g$-dimensional GD vectors (measured in seconds) of the two fiber types. For mode groups $i=1,2,\dots, N_g$,
\begin{align*}
    \tau_{1,i} = \frac{L_{1}}{d_i} \sum_{l\in\mathcal{M}_i}\beta^{(1)}_{1}[l],\\
   \tau_{2,i} = \frac{L_{2}}{d_i} \sum_{l\in\mathcal{M}_i}\beta^{(2)}_{1}[l],
\end{align*}
where $\beta^{(m)}_{1}[l],~m=1,2$ is the uncoupled GD (measured in seconds per meter) of the mode $l$ such that $\sum_{l=1}^{D}\beta^{(m)}_{1}[l]=0$, and $L_1$ and $L_2$ are the lengths of the fiber segments in one compensation period.
The $N_g\times N_g$ mode-group power coupling matrices of the mode permuter and mode scrambler are given by
\begin{align*}
    \PC[i,j] = \sum_{l\in\mathcal{M}_i}{\sum_{m \in\mathcal{M}_j}{ \left| \Rp[l,m] \right|^2}}, \\
    \SC[i,j] = \sum_{l\in\mathcal{M}_i}{\sum_{m \in\mathcal{M}_j}{ \left| \Rs[l,m] \right|^2}}, \\
\end{align*}
respectively. 
If mode permuters are not used in the link, $\PC$ can be set to $\D$, the diagonal matrix of mode group degeneracies. 
The GD STD after $K$ spans is given by
\begin{align}
\begin{split}
    \sigma_{\mathrm{GD}}(K) ={}&\sqrt{\frac{\E{\norm{\ttot}^2 }}{D}},
\end{split}
\label{eq:gd_rms_tot}
\end{align}
where $\E{\norm{\ttot}^2 }$ denotes the expected squared norm of the coupled GDs and is given by \eqref{eq:gd_exp_full}.
\begin{figure*}[!ht]
   \hrulefill
\begin{align}
    \begin{split}
    \E{\norm{\ttot}^2} =& K k_p\sum_{i=1}^{N_g} (\sigma_{\text{intra},1,i}^2 + \sigma_{\text{intra},2,i}^2 ) + Kk_p \left(\nu_1 \tautau_1^T \D \tautau_1 + \nu_2 \tautau_2^T \D \tautau_2 + \eta_1\eta_2 \tautau_1^T \PC \tautau_2 + \eta_1\eta_2 \tautau_2^T \PC \tautau_1 \right) \\
        & + \Biggl[2 \sum_{k=1}^{Kk_p-1} \Biggl( \eta_1^2 \tautau_1^T \left( \D \left( \Dinv \SC \Dinv \PC\right)^k + \D \left( \Dinv \PC \Dinv \SC\right)^k \right) \tautau_1 e^{\frac{2L_1}{\linter}} \\
        & \qquad\quad~+ \eta_2^2 \tautau_2^T \left( \D \left( \Dinv \SC \Dinv \PC\right)^k + \D \left( \Dinv \PC \Dinv \SC\right)^k \right) \tautau_2 e^{\frac{2L_2}{\linter}} \\
        & \qquad\quad~+ \eta_1\eta_2 \tautau_1^T \left( \PC \left( \Dinv \SC \Dinv \PC\right)^k + \left( \PC \left( \Dinv \SC \Dinv \PC\right)^k \right)^T \right) \tautau_2 \\
        & \qquad\quad~+ \eta_1\eta_2 \tautau_1^T \left( \D \left( \Dinv \SC \Dinv \PC\right)^{k-1}\Dinv \SC + \left(  \D \left( \Dinv \SC \Dinv \PC\right)^{k-1}\Dinv \SC \right)^T \right) \tautau_2 e^{\frac{2\lenS}{\linter}} \Biggr) \Biggr].
    \end{split}
    \label{eq:gd_exp_full}
\end{align}
\hrulefill
\end{figure*}
For fiber types $m=1,2$, $\sigma_{\textrm{intra},m,i} \propto \sqrt{L_m  L_{\textrm{intra},i}}$ is the intra-group GD STD of the $i$th mode group with intra-group coupling length $L_{\textrm{intra},i}$ over length $L_m$. 
The factors 
\begin{align}
    \eta_m &= \frac{\linter}{2L_m} \left (1-e^{-2L_m/\linter} \right ),\label{eq:eta}\\
    \nu_m &= \frac{\linter}{L_m} - \frac{\linter^2}{2L_m^2}\left (1-e^{-2L_m/\linter} \right ) \label{eq:nu}
\end{align}
are correction terms inspired by a Stokes-vector analysis incorporating finite characteristic length $\linter$ of random inter-group coupling \cite{vijay_arxiv_2025}.
$\linter$ is inversely related to the strength of random inter-group coupling.
In graded-index MMFs, there is strong random intra-group coupling and weak random inter-group coupling: $L_{\textrm{intra},i}\ll L_m, i=1,\dots,N_g$, and $L_m<\linter$.

While \eqref{eq:gd_exp_full} may appear complicated, a well-designed mode scrambler can make the terms in the square brackets negligible as $K$ increases \cite{vijay_modal_2025}.
Among the remaining terms, the intra-group GD spread terms $\sigma_{\text{intra},1,i}^2 + \sigma_{\text{intra},2,i}^2$ have negligible contributions as well, since $L_{\textrm{intra}}\ll \lenS$.
The key terms in \eqref{eq:gd_exp_full} are $(\nu_1 \tautau_1^T \D \tautau_1 + \nu_2 \tautau_2^T \D \tautau_2)$, the \textit{GD accumulation} term corresponding to the fiber segments individually, and $(\eta_1\eta_2 \tautau_1^T \PC \tautau_2 + \eta_1\eta_2 \tautau_2^T \PC \tautau_1)$, the \textit{GD compensation} term due to the correlated delays of the fiber segments.
The sum of these terms corresponds to the net effective GD accumulation in one span.
It is important to note that, owing to the inclusion of mode scramblers, a multiplicative factor \(K\) is involved.
While the accumulation terms are inherently positive, the compensation terms may assume negative values if the fibers and/or mode permuters are designed intentionally to achieve this effect.
Mathematically, we obtain the following inequalities:
\begin{align}
    |\tautau_1^T \D \tautau_1 + \tautau_2^T \D \tautau_2| &\geq | \tautau_1^T \PC \tautau_2 + \tautau_2^T \PC \tautau_1|, \label{eq:ineq1}\\
    \nu_1 + \nu_2 &\geq 2 \eta_1\eta_2 \label{eq:ineq2}.
\end{align}
Perfect compensation is possible only if both \eqref{eq:ineq1} and \eqref{eq:ineq2} are satisfied with equality. 
The first equality can be achieved if $\tautau_1, \tautau_2$ and (if applicable) $\PC$ are designed together.
The second equality is achieved only if $\linter\rightarrow\infty$.
This means that random inter-group coupling from the fiber segments is always unfavorable to GD compensation.

\subsection{Multi-Section Simulations}
In this sub-section, we verify the analytical expression for GD STD through multi-section simulations similar to those discussed in \cite{ho_linear_2014,vijay_effect_2023} for systems supporting three mode groups with a total of six spatial modes. 
We choose system configurations that achieve perfect GD compensation in the absence of random inter-group coupling in the fibers.
However, we evaluate the GD STD when random inter-group coupling is present.
The simulation parameters are provided in Table \ref{tab:siml_params} and the results are presented in Fig. \ref{fig:gd_theory}.

\begin{table}
    \setlength\tabcolsep{0pt}
    \centering
    \caption{Simulation Parameters I}
    \label{tab:siml_params}
    \begin{threeparttable}
        \begin{tabularx}{1\columnwidth} { >{\raggedright\arraybackslash}X  >{\centering\arraybackslash}X >{\raggedright\arraybackslash}X }
             \textbf{Parameter} & \textbf{Symbol}& \textbf{Value/Expression}  \\
             \hline \hline \\
             Number of spatial and polarization modes & $D$ & $12$\\
             Number of mode groups & $N_g$ & $3$\\
             Mode group degeneracy & $\mathbf{d}$ & $[2,4,6]^T$\\
             Number of spans  &$K$ & $100$\\
             Compensation periods per span  &$k_p$ & $1$\\
             Span length  &$\lenS$ & \\
             Characteristic inter-group coupling length  &$\linter$ & $\linter\in\{2\lenS, 20\lenS, 200\lenS\}$\\
             \hline\\
              \multicolumn{3}{c}{\textbf{\textit{Ideal Conventional Compensation}}}\\
                Mode-averaged GDs over one span & $(\tautau_1,\tautau_2)$ & $\tautau_1=-\tautau_2$\\
             Segment lengths &$(L_1,L_2)$ & $L_1\neq L_2$, e.g.~$\left( 0.6\lenS, 0.4\lenS \right)$\\
             \hline \\
             \multicolumn{3}{c}{\textbf{\textit{Ideal Self-Compensation}}}\\
                Mode-averaged GDs over one span & $(\tautau_1,\tautau_2)$ & $\tautau_1=\tautau_2$, $\tau_{1,1}=\tau_{1,2}=-\tau_{1,3}$\\
             Segment lengths &$(L_1,L_2)$ & $\left( 0.5\lenS, 0.5\lenS \right)$\\
             \hline \hline
        \end{tabularx}
            
    \end{threeparttable}
    
\end{table}

Conventional compensation and self-compensation can achieve perfect compensation of inter-group GD spread when $\linter \rightarrow \infty$ under the following conditions:
\begin{itemize}
    \item \textbf{\textit{Ideal conventional compensation}}: The delay vectors of the two fiber types are equal and opposite, $\tautau_1=-\tautau_2$.
            A mode permuter is not required between the two fiber segments and as a result, $\PC=\D$, as shown in Fig. \ref{fig:PC_ideal_comp}.
    \item \textbf{\textit{Ideal self compensation}}: Only one fiber type is used, $\tautau_2=\tautau_1$.
    For $N_g=3$, the average delays of the first two mode groups are equal in magnitude and sign. 
    The average delay of the third mode group is equal in magnitude but opposite in sign to that of the first two mode groups:
    \begin{equation*}
        \tau_{1,1}=\tau_{1,2}=-\tau_{1,3}.
    \end{equation*}
    
    A mode permuter that transfers all the power from the first two mode groups to the third mode group, and vice-versa, is needed between the two segments \cite{vijay_modal_2025}. It has a power coupling matrix
    \begin{equation*}
        \PC = \begin{bmatrix}
        0 & 0 & 2 \\
        0 & 0 & 4 \\
        2 & 4 & 0
    \end{bmatrix},
    \end{equation*}
    as shown in Fig. \ref{fig:PC_ideal_selfcomp}. For this choice of $\tautau_1$ and $\PC$, we find that $\tautau_1^T \D \tautau_1 + \tautau_1^T \PC \tautau_1 = 0$.
    It can be shown that this is the only configuration that achieves zero GD STD for MDM systems with mode-group degeneracies $\mathbf{d}=[2,4,6]^T$.
\end{itemize}

We investigate the effect of random inter-group coupling on GD compensation for $\linter\in\{2\lenS, 20\lenS, 200\lenS\}$.
Mode scramblers with dense power coupling matrices, as shown in Figs. \ref{fig:SC_ideal_comp} and \ref{fig:SC_ideal_selfcomp}, are used to reduce the accumulation of residual GD spread at the end of each span.

\begin{figure*}
\subfloat{\label{fig:PC_ideal_comp}}
\subfloat{\label{fig:SC_ideal_comp}}
\subfloat{\label{fig:GD_ideal_comp}}
\subfloat{\label{fig:PC_ideal_selfcomp}}
\subfloat{\label{fig:SC_ideal_selfcomp}}
\subfloat{\label{fig:GD_ideal_selfcomp}}
\centering
\includegraphics[width=0.49\textwidth]{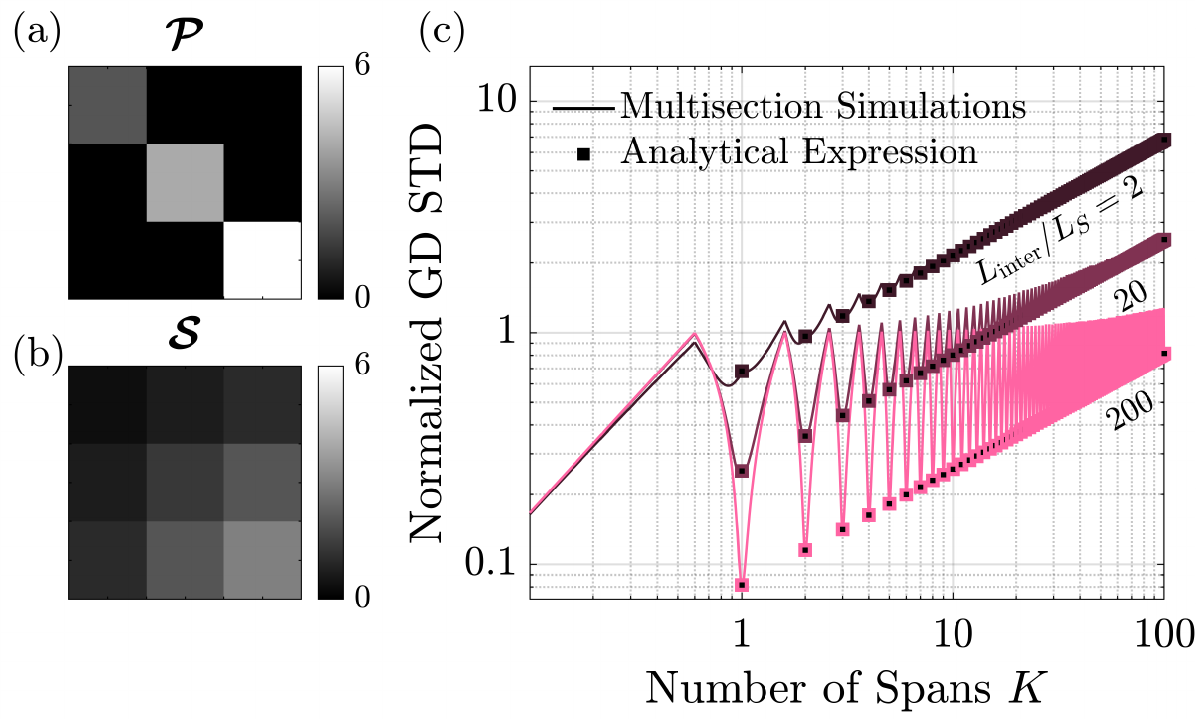}
\includegraphics[width=0.49\textwidth]{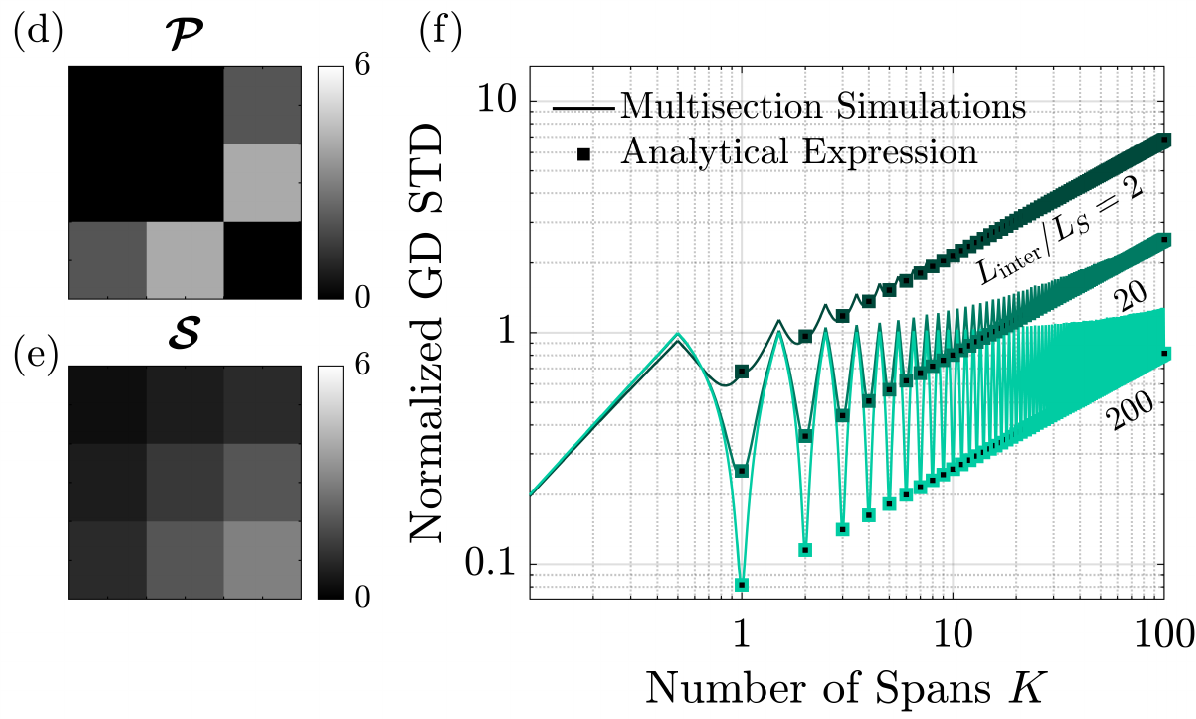}
\caption{Simulation of GD compensation for three values of inter-group coupling lengths for $k_p=1$ compensation period per span. (a – c) Conventional compensation; (d – f) Self compensation. 
(a, d) Mode permuter power coupling matrix; (b, e) Mode scrambler power coupling matrix; (c, f) GD STD as a function of number of spans. The solid lines correspond to numerical estimates from multi-section simulations and the square markers correspond to analytical estimates from the formula \eqref{eq:gd_rms_tot}. The GD STD is normalized to the one-segment GD STD.} 
\label{fig:gd_theory}
\end{figure*}

The GD STD is evaluated through multisection simulations after every fiber section and the analytical expression in \eqref{eq:gd_rms_tot} after every span. 
Figs. \ref{fig:GD_ideal_comp} and \ref{fig:GD_ideal_selfcomp} show the normalized GD STD as a function of the number of spans for ideal conventional compensation and self-compensation, respectively.
The zig-zag pattern illustrates the processes of GD accumulation and compensation. 
The plots indicate that compensation remains effective even with moderate random inter-group coupling. 
The estimates obtained from the analytical expressions agree closely with those derived from numerical simulations.
The overall trend follows a square-root relationship, which is expected owing to the use of mode scramblers. 
Notably, if the ratio $\linter/\lenS$ decreases by a factor of $10$, the asymptotic value of the GD STD increases by a factor of $\sqrt{10}$.

\subsection{System- and Fiber-Design Considerations}
\label{subsec:sys_fib_des}
From a system design perspective, to mitigate the effects of random inter-group coupling, the compensation scheme can be implemented multiple times within a single span.
For $k_{p}$ compensation periods per span, the GD STD will decrease by a factor of $\sqrt{k_{p}}$ in the absence of random inter-group coupling.
In terms of fiber design for GD compensation, it is essential to reduce the magnitudes of the uncoupled delays in the fiber segments. 
This is because the residual GD spread caused by random inter-group coupling is proportional to the GD accumulation terms.
If the uncoupled delays of the fiber segments are reduced by a factor $\kappa$, the GD STD will also decrease by the same factor of $\kappa$.
It is important to note that even when the length of inter-group coupling is similar to the span length, GD compensation can still be more effective in managing GD than no compensation at all for specific GD orderings.


\section{Design of Compensating Fibers Supporting Six Spatial Modes}
\label{sec:FiberDesign}
This section presents a design strategy for fibers that support six spatial modes, which is applicable to both conventional and self-compensating schemes. 
We provide design examples and evaluate system performance across the C-band (1530 nm to 1565 nm). 
Our primary goal is to minimize GD STD, while also ensuring that the designs are tolerant to random inter-group coupling and errors in fiber fabrication. 
Additionally, the designs must meet specifications for macro-bending loss and cut-off wavelength \cite{noauthor_transmission_nodate}.
The design strategy involves selecting a cost function and a search region that take all these factors into account. 
We parameterize the fiber design to facilitate numerical optimization techniques such as particle swarm optimization since the cost function is non-convex over the search space.

\subsection{Cost Functions for Optimal GD Compensation}
We formulate minimization problems for conventional and self-compensation.
For conventional compensation, we minimize a wavelength-averaged cost function $J^{(\textrm{conv})}(\lambda_i)$, which is a function of $n^{(1)}, n^{(2)}$, the refractive index profiles  of the two fiber types, and $f = L_1/(L_1+L_2)$, the ratio of the length of the first fiber to the combined length:
\begin{equation}
    \label{eq:optComp}
    \begin{aligned}
        \minimize_{n^{(1)}, n^{(2)}, f}     &\quad \frac{1}{N_\lambda}\sum_{i=1}^{N_\lambda} J^{(\textrm{conv})}(\lambda_i),\\
        \subjectto          &\quad 1/3\leq f \leq 2/3.
    \end{aligned}\quad
\end{equation}
We limit the ratio $f$ to avoid large disparities in segment lengths.
Similarly, for self-compensation, we minimize a wavelength-averaged cost function $J^{(\textrm{self})}(\lambda_i)$ which is a function of $n$, the refractive index profile of the self-compensation fiber:
\begin{equation}
    \label{eq:optSelfComp}
    \begin{aligned}
        \minimize_{n}     &\quad \frac{1}{N_\lambda}\sum_{i=1}^{N_\lambda} J^{(\textrm{self})}(\lambda_i),\\
    \end{aligned}\quad
\end{equation}

The cost functions are given by
\begin{align}
\label{eq:cost_reg}
    J^{(\textrm{conv})}(\lambda)=& \norm{f \barbeta_1^{(1)} + (1-f) \barbeta_1^{(2)} } \nonumber\\
    &+ \gamma_{\textrm{CD}} \norm{f \barCD^{(1)} + (1-f) \barCD^{(2)} } \nonumber\\
    &+ \gamma_{\textrm{GD}} \left( \norm{f\bbeta_1^{(1)}} + \norm{f\bbeta_1^{(2)}} \right),
\end{align}
where, for fiber types $m=1,2$, $\bbeta_1^{(m)}$ is the vector of uncoupled modal GDs, $\barbeta_1^{(m)},m=1,2$ is the vector of mode-group-averaged uncoupled GDs, and $\barCD^{(m)}$ is the vector of mode-group-averaged CD parameters, and
\begin{align}
    \label{eq:cost_self}
    J^{(\textrm{self})}(\lambda)=& \biggl \{2(\beta_1[01] - \beta_1[11])^2 + (\beta_1[01] + \beta_1[02])^2 + \nonumber\\
    &\quad2(\beta_1[01] + \beta_1[21])^2 + 2(\beta_1[02] - \beta_1[21])^2 \biggr\}^{1/2} \nonumber\\
    &+ \gamma_{\textrm{CD}} \norm{\barCD} \nonumber\\
    &+ \gamma_{\textrm{GD}} \left( \norm{g_{\textrm{penalty}}(\bbeta_1,\beta_{th})} \right),
\end{align}
where $\bbeta_1$ is the vector of uncoupled modal GDs and $\beta_1[l\,m]$ represents the uncoupled GD of spatial mode $\LP_{l\,m}$, $\barCD$ is the vector of mode-group-averaged CD parameters, and $g_{\textrm{penalty}}$ is a regularization term defined for a threshold $\beta_{th}$ as
\begin{equation*}
    g_{\textrm{penalty}}(x,\beta_{th}) = \begin{cases} 
    \beta_{th} \sqrt{1 - x/\beta_{th}}, &x<\beta_{th} \\
    x - \beta_{th}, & x\geq \beta_{th}.        
    \end{cases}
\end{equation*}
Both cost functions \eqref{eq:cost_reg} and \eqref{eq:cost_self} have three components: the first captures the effective GD spread, 
the second captures the effective MDCD spread and is multiplied by a hyperparameter $\gamma_{\textrm{CD}}$, 
and the third captures the STD of the uncoupled GDs and is multiplied by a hyperparameter $\gamma_{\textrm{GD}}$.

The CD regularization term in each cost function is included in addition to the wavelength averaging in \eqref{eq:optComp} and \eqref{eq:optSelfComp} to ensure wavelength-flat GD compensation performance.
The GD regularization term in \eqref{eq:cost_reg} is to ensure low uncoupled GD STD to reduce the effect of random inter-group coupling.
The GD regularization term in \eqref{eq:cost_self} performs a similar function, except it also penalizes designs with very low uncoupled GDs, which can affect fabrication tolerances (discussed in Section \ref{subsec:errors}).

\subsection{Parameterized Fiber Profile}
A parametric approach to fiber design facilitates the use of numerical optimization methods, such as particle-swarm optimization, which are preferred since the cost function is a non-convex function of the refractive index profile. 
Previous research has suggested parameterized fiber design for GD management and compensation \cite{gruner-nielsen_few_2012, mori_low_2013, mori_few-mode_2014}. 
Parameters defining the fiber profile, including core-to-cladding index contrast, core radius, and the $\alpha$-parameter describing the core's gradation, significantly influence propagation constants and modal dispersion.
\begin{figure}
\centering
\includegraphics[width=0.8\columnwidth]{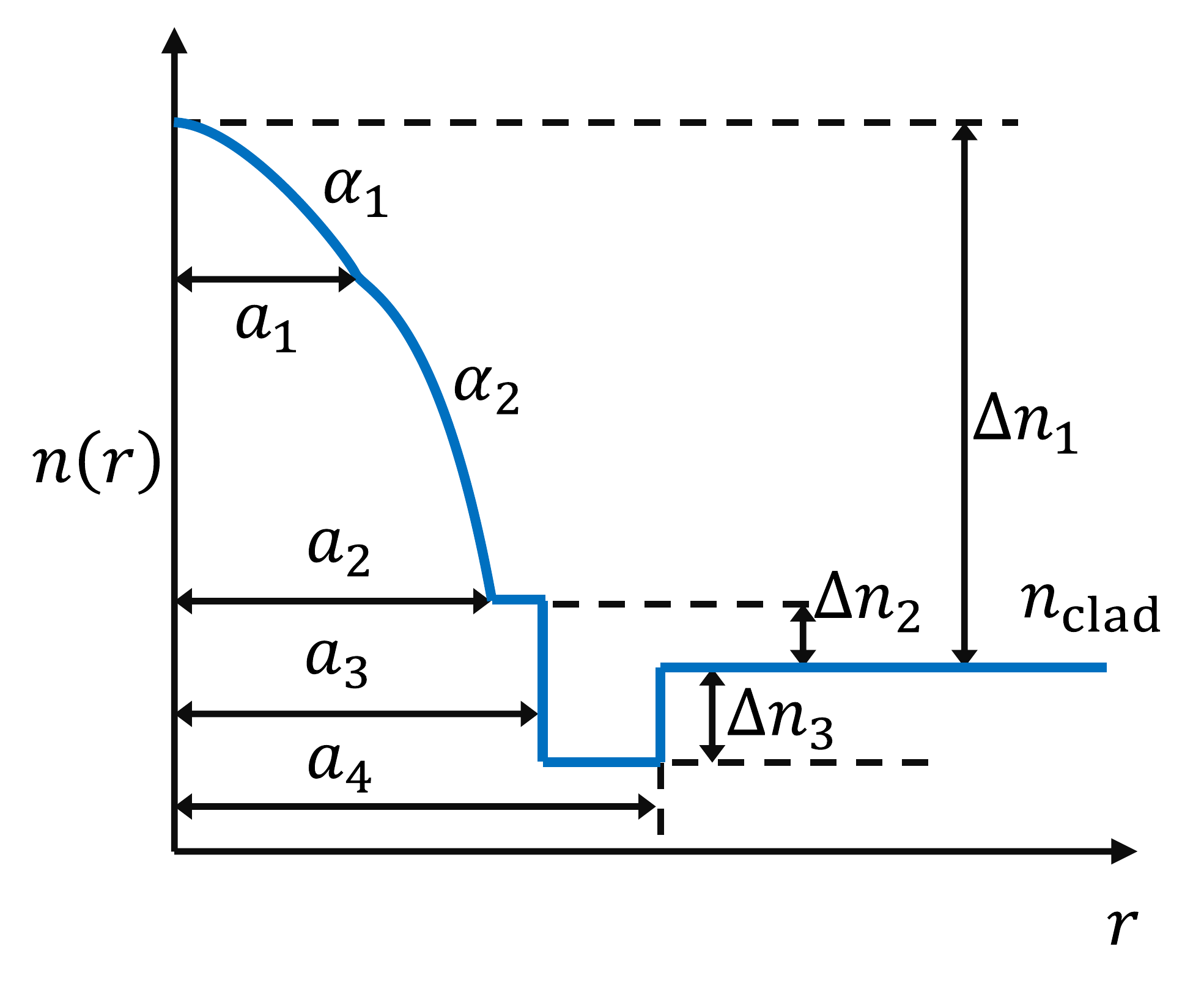}
\caption{Refractive index profile of a graded-index, trench-assisted compensating fiber. The inner and outer parts of the core have (possibly) different $\alpha$-law gradations. There is a pedestal/cliff feature at the core-cladding boundary.}
\label{fig:ri_profile}
\end{figure}

Fig. \ref{fig:ri_profile} plots the refractive index profile highlighting the important features and parameters.
We define a refractive index profile with nine parameters as follows:
\begin{align}
    n(r) = \begin{cases}
        (\nclad + \Delta n_2) \sqrt{\frac{1 - 2 \Delta_1 \left(\frac{r}{a_1} \right)^{\alpha_1}}{1 - 2\Delta_1}}, & 0\leq r\leq a_1 \\
        (\nclad + \Delta n_2) \sqrt{\frac{1 - 2 \Delta_2 \left(\frac{r}{a_2} \right)^{\alpha_2}}{1 - 2\Delta_2}}, & a_1< r\leq a_2 \\
        \nclad + \Delta n_2, & a_2< r\leq a_3 \\
        \nclad - \Delta n_3, & a_3< r\leq a_4 \\
        \nclad, & a_4<r
    \end{cases},
    \label{eq:ref_ind}
\end{align}
where $r$ is the radial position, and
\begin{align}
    \Delta_1 &= \frac{1}{2}\left( 1 - \left(\frac{\nclad + \Delta n_2}{\nclad + \Delta n_1}\right)^2 \right), \\
    \Delta_2 &=  \frac{\Delta_1 \left(1 - \left(\frac{a1}{a_2} \right)^{\alpha_1} \right)}{1 - \left(\frac{a1}{a_2} \right)^{\alpha_2} + 2\Delta_1 \left(\left(\frac{a1}{a_2} \right)^{\alpha_2} - \left(\frac{a1}{a_2} \right)^{\alpha_1} \right)}.
\end{align}
The pedestal/cliff feature at the core-cladding boundary can be adjusted to control the CD behavior.
The different $\alpha$ parameters for the inner and outer core provide control of uncoupled GDs in different mode groups.

\subsection{Design Examples and System Performance}
\begin{table*}
    \setlength\tabcolsep{0pt}
    \centering
    \caption{Fiber Design Parameters}
    \label{tab:design_params}
    \begin{threeparttable}
        \begin{tabularx}{1\linewidth} {>{\raggedright\arraybackslash}X >{\centering\arraybackslash}X  >{\centering\arraybackslash}X >{\centering\arraybackslash}X  >{\centering\arraybackslash}X  >{\centering\arraybackslash}X >{\centering\arraybackslash}X  >{\centering\arraybackslash}X  >{\centering\arraybackslash}X >{\centering\arraybackslash}X >{\raggedright\arraybackslash}X }%
              & $\alpha_1$ & $\alpha_2$ & $\Delta n_1$ & $\Delta n_2$ & $\Delta n_3$ & $a_1 (\mu\textrm{m})$ & $a_2 (\mu\textrm{m})$ & $a_3 (\mu\textrm{m})$ & $a_4 (\mu\textrm{m})$  & Notes \\
             \hline \hline \\
             \multicolumn{11}{c}{\textbf{\textit{Conventional Compensation}}}\\
            Fiber 1 & $1.962$ & $1.962$ & $8.92\cdot 10^{-3}$ & $1.21\cdot 10^{-3}$ & $3.15\cdot 10^{-3}$ & $5.7$ & $11.38$ & $12.52$ & $19.41$ & $f=1/3$\\
            Fiber 2 & $1.989$ & $1.989$ & $7.45\cdot 10^{-3}$ & $1.43\cdot 10^{-3}$ & $3.11\cdot 10^{-3}$ & $5.89$ & $11.77$ & $13.4$ & $19.54$ & $(1-f)=2/3$\\
            \hline\\
            \multicolumn{11}{c}{\textbf{\textit{Self-Compensation}}}\\
            Fiber 1 & $1.925$ & $1.994$ & $7.43\cdot 10^{-3}$ & $1.46\cdot 10^{-3}$ & $3.03\cdot 10^{-3}$ & $1.66$ & $11.77$ & $14.6$ & $19.78$ & \\ \\
             \hline \hline
        \end{tabularx}
            
    \end{threeparttable}
\end{table*}

\begin{figure}
\centering
\includegraphics[width=\columnwidth]{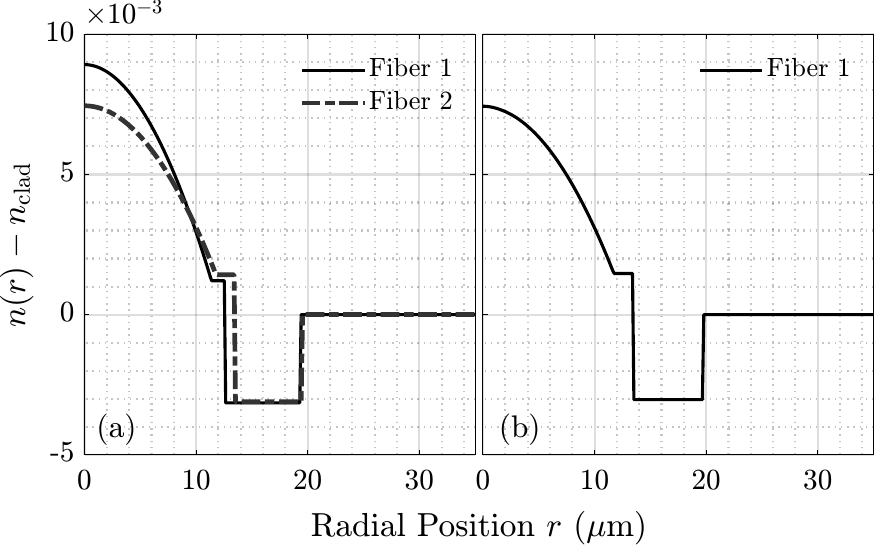}
\caption{Design example for compensation. Optimized fiber refractive index profiles relative to the cladding index for (a) conventional compensation and (b) self-compensation.}
\label{fig:sim_ri_profile}
\end{figure}

We perform particle-swarm optimization \cite{kennedy_particle_1995} to search for an optimal pair of fibers for conventional compensation \eqref{eq:optComp} and an optimal fiber for self-compensation \eqref{eq:optSelfComp}.
The hyper-parameters for conventional compensation are chosen to be $\gamma_{\textrm{CD}}=0.05, \gamma_{\textrm{GD}}=0.005$, and those for self-compensation are chosen to be $\gamma_{\textrm{CD}}=1, \gamma_{\textrm{GD}}=0.1, \beta_{th}=10\,\textrm{ps/km}$.
The parameter search space is chosen such that the fibers support six spatial modes in each polarization, and comply with leaky- and guided-mode bending loss standards over the C-band. 
The refractive index contrast between the center of the core and the trench is consistent with a germanium-free core, enabling low transmission loss \cite{sillard_germanium-free_2024, sillard_germanium-free_2025}.

\subsubsection{Fiber Design Examples}
Table \ref{tab:design_params} provides the parameters of the optimized fibers.
Fig. \ref{fig:sim_ri_profile} shows the refractive index profiles relative to the cladding as a function of the radial position for the conventional compensation fibers and the self-compensation fiber.
Fig. \ref{fig:gd_fiber} shows the dispersion performance of the optimized fibers over the C-band.
Here are the key observations.
\paragraph*{Conventional Compensation}
\begin{itemize}
    \item Fig. \ref{fig:scaledGD_comp} shows the scaled uncoupled GDs of the spatial modes of the two fibers, $f\bbeta_1^{(1)}$ and $-(1-f)\bbeta_1^{(2)}$, as a function of wavelength.
    The scaled delays align closely, indicating good optimization over the entire C-band.
    \item Fig. \ref{fig:CDeff_comp} shows the effective CD parameters as a function of wavelength.
    The effective CD parameter $f \textrm{\textbf{CD}}^{(1)} + (1-f) \textrm{\textbf{CD}}^{(2)}$ is the same for each mode.
    Consequently, the effective uncoupled GD STD exhibits a flat wavelength trend in the C-band, as seen in Fig. \ref{fig:GDeff_comp}.
\end{itemize}
\paragraph*{Self-Compensation}
\begin{itemize}
    \item Fig. \ref{fig:scaledGD_self_comp} shows the uncoupled GDs of the spatial modes of the optimized fiber, $\bbeta_1$ as a function of wavelength.
    As desired, the delays of the first two mode groups with spatial modes $\{\LP_{01}\},\{\LP_{11a},\LP_{11b}\}$ are almost equal.
    Their average is equal to the negative of the average of the delays of the third mode group with spatial modes $\{\LP_{02},\LP_{21a},\LP_{21b}\}$.
    We also observe that the mode-group-averaged GDs are not too close to zero, which would negatively impact fabrication error tolerances.
    \item Fig. \ref{fig:CDeff_self_comp} shows the uncoupled CD parameters of the spatial modes $\textrm{\textbf{CD}}$ as a function of wavelength.
    The CD is the same for each mode.
    \item An effective uncoupled GD STD can be defined as $\mathrm{std}(0.5\bbeta_1+0.5\bbeta_1^{(\textrm{perm})})$, where $\bbeta_1^{(\textrm{perm})}$ represents the uncoupled GD vector after mode permutations.
    In this case, the mode permutations exchange power between $\LP_{01}$ and $\LP_{02}$ and between $\LP_{11}$ and $\LP_{21}$.
    As a result of low MDCD and effective optimization, the effective GD STD is small and exhibits a flat wavelength trend in the C-band, as seen in Fig. \ref{fig:GDeff_self_comp}.
\end{itemize}

Fig. \ref{fig:gd_fiber} is intended to characterize the uncoupled dispersion properties of the fibers, and does not take into account random intra- and inter-group coupling when calculating dispersion performance. The statistics of the coupled GDs are highly relevant to the complexity of DSP at the receiver, and are studied via multi-section simulations in the following subsection. 

\begin{figure}
\centering
\subfloat{\label{fig:scaledGD_comp}}
\subfloat{\label{fig:CDeff_comp}}
\subfloat{\label{fig:GDeff_comp}}
\subfloat{\label{fig:scaledGD_self_comp}}
\subfloat{\label{fig:CDeff_self_comp}}
\subfloat{\label{fig:GDeff_self_comp}}
\includegraphics[width=\columnwidth]{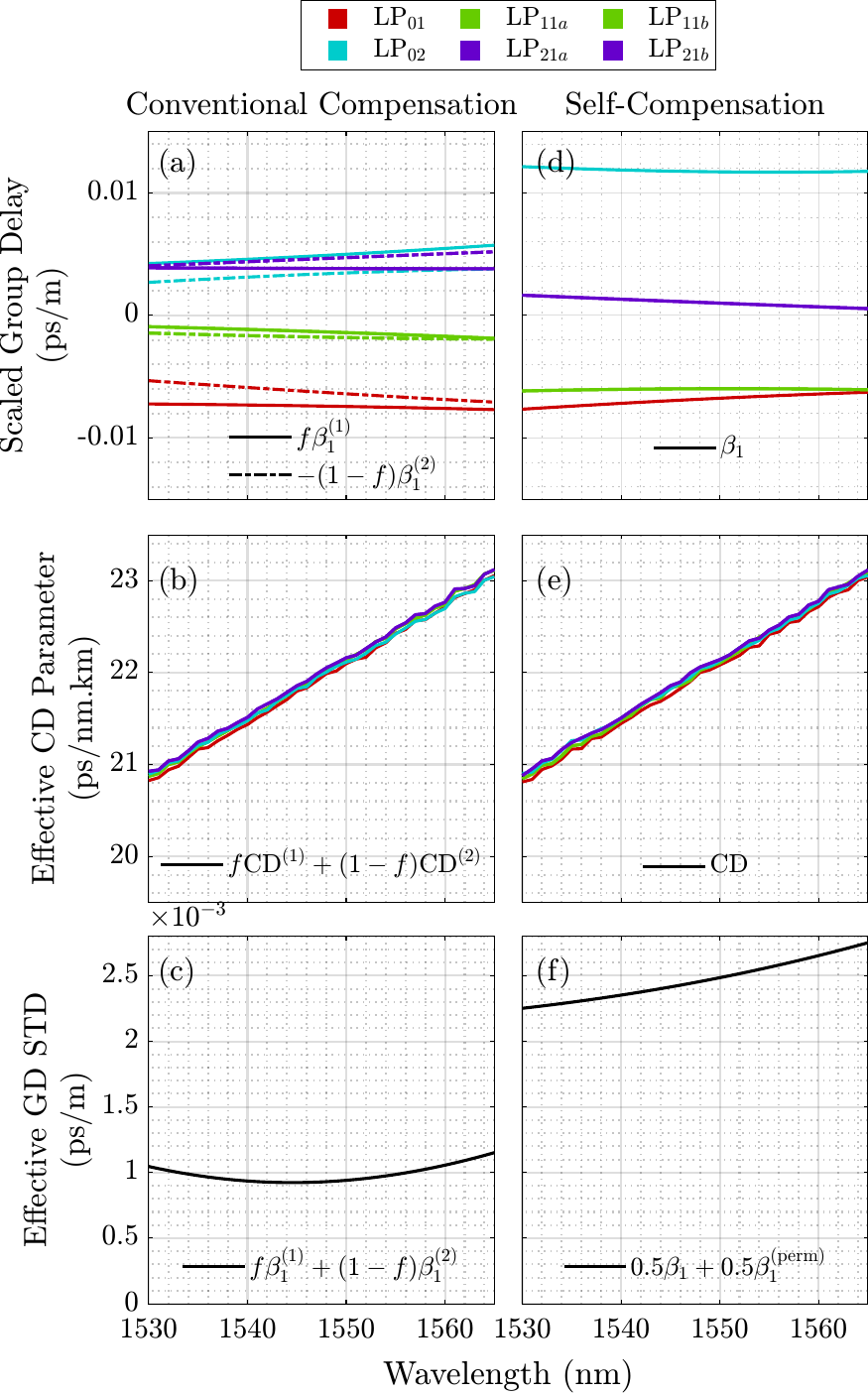}
\caption{Design example for compensation. (a - c) Conventional compensation; (d - f) Self-compensation. (a, d) Scaled uncoupled GDs as a function of wavelength over the C-band of the spatial modes of the optimized fiber(s); (b, e) Effective uncoupled CD parameter as a function of wavelength over the C-band; (c, f) Effective uncoupled GD STD as a function of wavelength over the C-band.}
\label{fig:gd_fiber}
\end{figure}

\subsubsection{Multi-Section Simulations}
We perform numerical multisection simulations to calculate the end-to-end GD STD for the optimized fibers. 
We select a span length $\lenS=50~\textrm{km}$.
We assume strong random intra-group coupling and weak random inter-group coupling in the fiber segments \cite{ho_linear_2014}.
\begin{figure*}
\subfloat{\label{fig:sys_a}}
\subfloat{\label{fig:sys_b}}
\subfloat{\label{fig:sys_c}}
\subfloat{\label{fig:sys_d}}
\subfloat{\label{fig:sys_e}}
\subfloat{\label{fig:sys_f}}
\subfloat{\label{fig:sys_g}}
\subfloat{\label{fig:sys_h}}
\centering
\includegraphics[width=\textwidth]{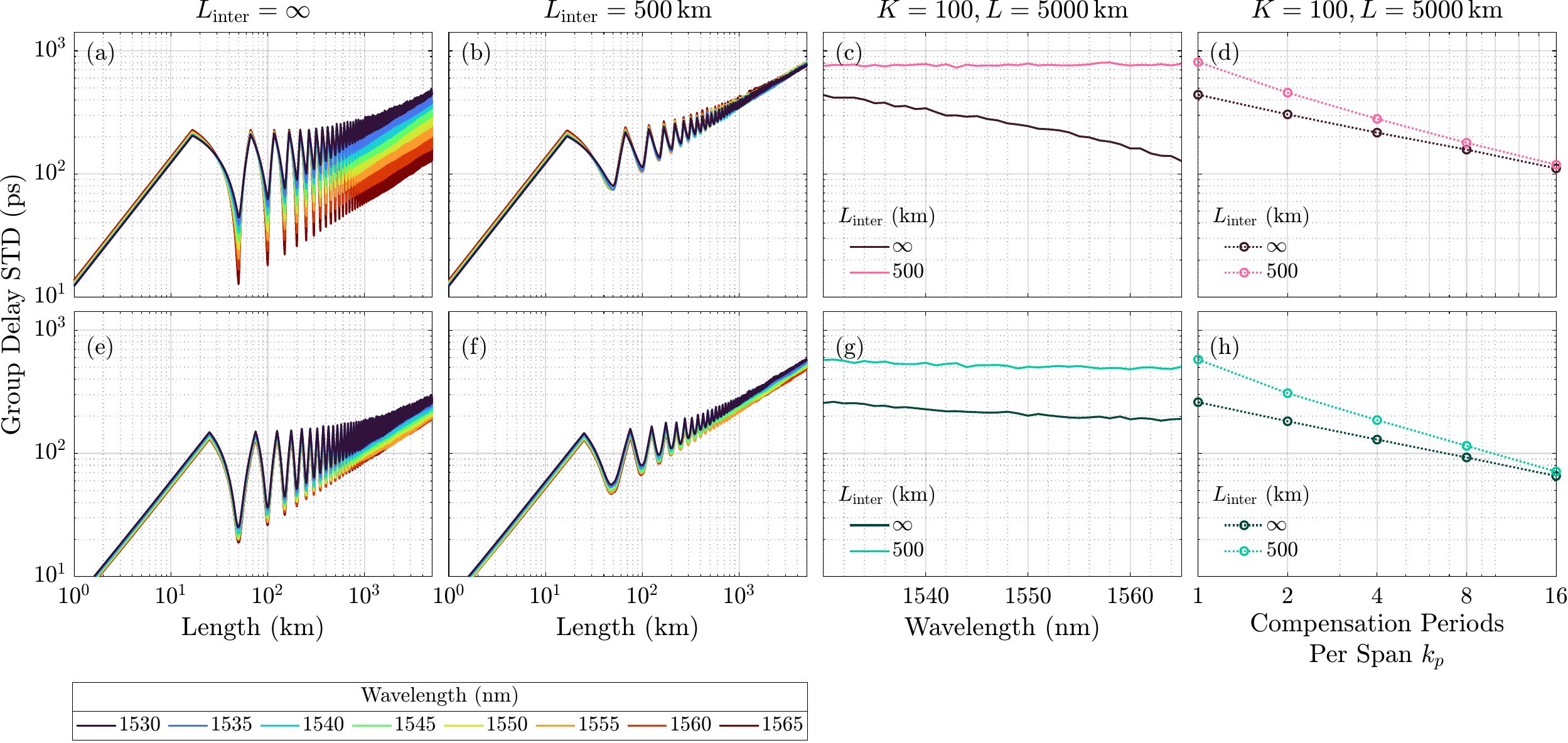}
\caption{
Simulation of compensation. (a - d) Conventional compensation; (e - h) Self-compensation. (a, e) GD STD as a function of length for various wavelengths in the C-band when $\linter = \infty$ and $k_p=1$; (b, f) GD STD as a function of length for various wavelengths in the C-band when $\linter = 500\,\text{km}$ and $k_p=1$; (c, g) GD STD as a function of wavelength after propagation over 5000 km for $\linter = \infty, 500\,\text{km}$ and $k_p=1$; (d, h) GD STD as a function of number of compensation periods per span $k_p$ after propagation over 5000 km for $\linter = \infty, 500\,\text{km}$.
}
\label{fig:gd_sys_sim}
\end{figure*}

Random inter-group coupling is influenced by various factors, including effective refractive index differences between mode groups and manufacturing processes involved.
The literature reports a wide range of coupling strengths. 
For instance, one study reports coupling strengths of less than $-20~\textrm{dB}$ over $10.5~\textrm{km}$ of a GD compensation system \cite{mori_few-mode_2014} where the fibers support six spatial modes and have index contrast similar to the designs presented in this paper.  
The reported coupling strength corresponds to a characteristic coupling length exceeding $700~\textrm{km}$.
In our simulations, we set the value of the inter-group coupling length to $\linter = 500~\textrm{km}$, which is equal to ten span lengths.
Additionally, we compare our results to a scenario with no random inter-group coupling, represented by $\linter=\infty$.
For the conventional compensation system, we set the lengths of the fiber segments according to the optimized length fractions: $L_1=f\lenS/k_p$, $L_2=(1-f)\lenS/k_p$.
For the self-compensation system, we set the lengths of the fiber segments to be equal and assume ideal mode permutations.

Fig. \ref{fig:gd_sys_sim} shows the GD STD as a function of the length of the link over $K=100$ spans for several wavelength channels in the C-band.
For the conventional compensation system, for $k_p=1$, the worst-case GD STD after $K=100$ spans is $438~\textrm{ps}$, effectively $6.2~\textrm{ps}/\sqrt{\textrm{km}}$, when $\linter=\infty$, as seen in Fig. \ref{fig:sys_a}, and $803~\textrm{ps}$, effectively $11.4~\textrm{ps}/\sqrt{\textrm{km}}$, when $\linter=500~\textrm{km}$, as seen in Fig. \ref{fig:sys_b}.
For the self-compensation system, for $k_p=1$, the worst-case GD STD after $K=100$ spans is $257~\textrm{ps}$, effectively $3.6~\textrm{ps}/\sqrt{\textrm{km}}$, when $\linter=\infty$, as seen in Fig. \ref{fig:sys_e}, and $575~\textrm{ps}$, effectively $8.1~\textrm{ps}/\sqrt{\textrm{km}}$, when $\linter=500~\textrm{km}$, as seen in Fig. \ref{fig:sys_f}.

Figs. \ref{fig:sys_d} and \ref{fig:sys_h} illustrate the impact of increasing the number of compensation periods per span $k_p$. As discussed in Section \ref{subsec:sys_fib_des}, GD STD decreases with the square root of $k_p$, provided there is no inter-group coupling. 
This reduction in GD STD is more pronounced in the presence of inter-group coupling.
For the conventional compensation system, as $k_p$ increases from $1$ to $16$, the worst-case GD STD reduces from $438~\textrm{ps}$ to $109~\textrm{ps}$ ($4.9~\textrm{ps}/\sqrt{\textrm{km}}$) when $\linter=\infty$, and from $803~\textrm{ps}$ to $115~\textrm{ps}$ ($5.1~\textrm{ps}/\sqrt{\textrm{km}}$) when $\linter=500~\textrm{km}$.
For the self-compensation system, the worst-case GD STD reduces from $257~\textrm{ps}$ to $65~\textrm{ps}$ ($2.9~\textrm{ps}/\sqrt{\textrm{km}}$) when $\linter=\infty$, and from $575~\textrm{ps}$ to $70~\textrm{ps}$ ($3.1~\textrm{ps}/\sqrt{\textrm{km}}$) when $\linter=500~\textrm{km}$.
When $k_p$ is very large, the effect of inter-group coupling becomes negligible.
However, this choice has its drawbacks. 
The number of splices, mode scramblers, and permuters required is proportional to $k_p$, resulting in higher losses. 
Consequently, the mode-averaged loss will increase linearly, and the MDL STD will increase with the square root of $k_p$.
Overall, the results indicate that effective group delay compensation is achievable over the C-band with thoughtful fiber and system design; however, there are trade-offs to consider.

The effective GD STD of the design examples in this paper is comparable to those reported for coupled MCFs supporting a similar number of modes \cite{saitoh_multi-core_2022, hayashi_spatial_2019, sakuma_microbending_2019}.
Optimized three-core coupled MCFs have achieved GD STD as low as $4.2~\textrm{ps}/\sqrt{\textrm{km}}$ \cite{hayashi_spatial_2019}.
Additionally, four-core coupled MCFs have reported a GD STD of $3.14~\textrm{ps}/\sqrt{\textrm{km}}$ with an optimized fiber bend radius of 31 cm, and $10.6~\textrm{ps}/\sqrt{\textrm{km}}$ in loosely bent cables \cite{sakuma_microbending_2019}. 
One important issue in such coupled MCF systems is their sensitivity to the bend radius. 
Tight or excessively loose bends in the deployed fibers can lead to an increase in GD STD \cite{sakuma_microbending_2019}.
Additionally, fabrication errors can impact the core radius and refractive index, which also affects the GD STD \cite{saitoh_multi-core_2022}. 
Fabrication errors are also an issue in GD-compensating MMF systems, as discussed in Section \ref{subsec:errors}.

\section{Discussion}
\label{sec:Discussion}
In this section, we discuss additional aspects of GD compensation.
\subsection{Realization of Mode Permuters}
GD self-compensation requires devices capable of performing precise mode permutations. One way to achieve these mode permutations is by utilizing a fan-out and fan-in configuration \cite{shibahara_long-haul_2020,di_sciullo_reduction_2023}. 
In this configuration, a mode demultiplexer directs the signals from the spatial modes of the multimode fiber (MMF) into individual single-mode fibers (SMFs). 
Subsequently, a mode multiplexer combines the signals back into the MMF after executing the desired permutation.
However, this approach presents several challenges, including high mode-averaged and mode-dependent losses, as well as undesirable modal crosstalk. 

An alternative mode permuter design uses a cascade of several long-period fiber grating (LPFG)-based mode couplers \cite{krutko_low-loss_2025}. The design of each LPFG, including its grating period and symmetry, is intended to execute a subset of the necessary mode permutations. Undesired couplings by the LPFGs are suppressed by propagation constant engineering, in which the transverse refractive index profile is optimized to render the propagation constant spacings for the undesired couplings significantly different from those for the desired couplings. The LPFG-based permuter design promises to achieve much lower mode-averaged and mode-dependent losses and modal crosstalk than designs based on fan-out and fan-in.

\subsection{Sensitivity to Fiber Fabrication Errors}
\label{subsec:errors}
Changes to the designed refractive index profile of fibers can impact the modal propagation constants, which in turn affect the modal GDs. 
Variations in the uncoupled modal GDs of the compensating fibers can result in an increased effective GD STD. 
In extreme cases, these changes can alter the order of the modal GDs, potentially rendering a compensation scheme completely ineffective.
We study the sensitivity of GD compensation to fiber fabrication errors by introducing Gaussian random noise to the refractive index profile of the fiber, similar to previous studies \cite{gololobov_multimode_2023,krutko_ultra-low-loss_2024}.
Research shows that different pieces of fiber drawn from the same preform can exhibit different refractive index errors \cite{gruner-nielsen_few_2012}.
Therefore, in our analysis, we assume that each fiber segment has an independent realization of refractive index error. 

\subsubsection*{GD STD Penalty}
In the presence of refractive index errors, the expression for the effective GD STD can be modified to include a penalty term:
\begin{equation}
\label{eq:err_GDS}
    \sigma_{\mathrm{GD}}(K) =\sqrt{\sigma_{\mathrm{GD},\textrm{ideal}}^2 (K) + \sigma_{\mathrm{GD},\textrm{err}}^2 (K) },
\end{equation}
where $\sigma_{\mathrm{GD},\textrm{ideal}}$ is given by the expression in \eqref{eq:gd_exp_full} and $\sigma_{\mathrm{GD},\textrm{err}}$ is the contribution of the refractive index errors. Using first-order perturbation theory, we can estimate the deviation in the GDs resulting from refractive index profile errors.
Appendix \ref{app:error_derivation} provides the derivation of this relationship.
The penalty due to profile errors is given by:
\begin{equation}
    \begin{split}
    \sigma_{\mathrm{GD},\textrm{err}}^2 (K) =& \left(K\frac{\nu_1 L_1 + \nu_2 L_2}{D} \right)\\
    &\times\tr{\left ( \D - \frac{\mathbf{d}\mathbf{d}^T}{D}  \right )\Sigma_n} \sigDeln^2,
\end{split}
\end{equation}
where $\sigDeln$ is the STD of the refractive index errors, and $\Sigma_n \sigDeln^2$ is the $N_g\times N_g$ covariance matrix of mode-group-averaged GD deviations.
For the design examples in this paper,
\begin{equation}
    \Sigma_n = 10^{-18}\begin{bmatrix}
        0.40&    0.29&    0.11 \\
        0.29&    0.47&    0.33 \\
        0.11&    0.33&    0.50
    \end{bmatrix} (\mathrm{s}^2/\mathrm{m}).
\end{equation}
We observe that $\Sigma_n$ shows minimal variation with wavelength across the C-band.

\begin{figure*}
\subfloat{\label{fig:err_a}}
\subfloat{\label{fig:err_b}}
\subfloat{\label{fig:err_c}}
\subfloat{\label{fig:err_d}}
\centering
\includegraphics[width=\textwidth]{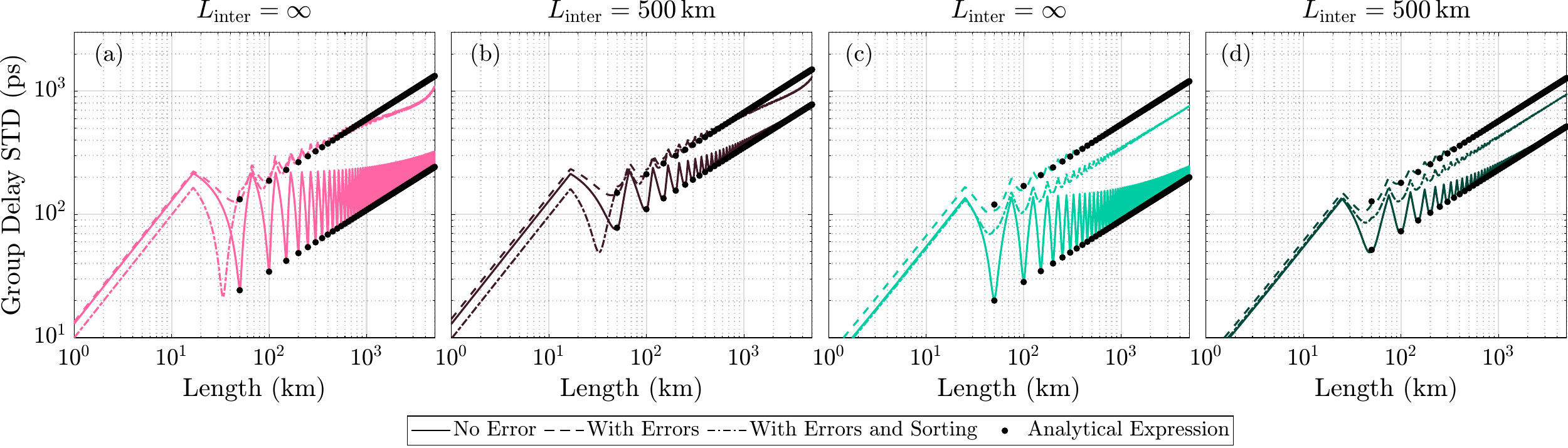}
\caption{Compensation in the presence of refractive index errors for $k_p=1$. (a - b) Conventional compensation; (c - d) Self-compensation. (a, c) GD STD as a function of length at 1550 nm for $\linter =\infty$; (b, d) GD STD as a function of length at 1550 nm for $\linter = 500\,\text{km}$; 
The solid lines correspond to GD performance with no refractive index errors, $\sigDeln=0$; 
the dashed lines correspond to the GD performance with refractive index errors of STD $\sigDeln=10^{-5}$; 
the dotted-dashed lines correspond to the GD performance with fiber sorting and refractive index errors of STD $\sigDeln=10^{-5}$;
and the solid circles correspond to analytical estimates.
}
\label{fig:gd_err_sim}
\end{figure*}
\subsubsection*{Multi-Section Simulations}
We perform multisection simulations for the design examples presented in Section \ref{sec:FiberDesign} for $k_p=1$.
Fig. \ref{fig:gd_err_sim} plots the GD STD for conventional and self-compensation for inter-group coupling lengths, $\linter = \infty$ and $\linter = 500 \textrm{ km}$ for the following cases:
\begin{itemize}
    \item No refractive index error
    \item With refractive index error of STD $\sigDeln=10^{-5}$.
    \item With refractive index error of STD $\sigDeln=10^{-5}$ and fiber sorting.
\end{itemize}
Fiber sorting involves characterizing the fabricated fibers. For each selected fiber from the set of fabricated fibers, another one that minimizes the effective GD STD of the combination is chosen as its pair. This process is repeated without replacement until all fibers are paired.
The plots also include the estimates from the analytical expression in \eqref{eq:err_GDS}.
The expectation in the GD variance is computed over the realizations of the random coupling matrices and refractive index errors.

From Fig. \ref{fig:gd_err_sim}, we observe that the analytical expression accurately captures the GD STD penalty caused by refractive index errors. 
For the conventional compensation system, the worst-case GD STD after $K = 100$ spans is $1.33~\textrm{ns}$ ($18.8~\textrm{ps}/\sqrt{\textrm{km}}$), when $\linter = \infty$, as shown in Fig. \ref{fig:err_a}, and increases to $1.50~\textrm{ns}$ ($21.2~\textrm{ps}/\sqrt{\textrm{km}}$) when $\linter = 500~\textrm{km}$, as seen in Fig. \ref{fig:err_b}. 
Fiber sorting provides only marginal improvement, resulting in a GD STD of $1.09~\textrm{ns}$ ($15.4~\textrm{ps}/\sqrt{\textrm{km}}$) for $\linter = \infty$ and $1.28~\textrm{ns}$ ($18.1~\textrm{ps}/\sqrt{\textrm{km}}$) for $\linter = 500~\textrm{km}$.  

For the self-compensation system, the worst-case GD STD after $K = 100$ spans is $1.20~\textrm{ns}$ ($17~\textrm{ps}/\sqrt{\textrm{km}}$) when $\linter = \infty$, as shown in Fig. \ref{fig:err_c}, and $1.28~\textrm{ns}$ ($18.1~\textrm{ps}/\sqrt{\textrm{km}}$) when $\linter = 500~\textrm{km}$, as seen in Fig. \ref{fig:err_d}. 
In this case, fiber sorting proves to be more effective, reducing the GD STD to $0.76~\textrm{ns}$ ($10.7~\textrm{ps}/\sqrt{\textrm{km}}$) for $\linter = \infty$ and $0.94~\textrm{ns}$ ($13.3~\textrm{ps}/\sqrt{\textrm{km}}$) for $\linter = 500~\textrm{km}$.  

After just one span, the effective GD STD is significantly higher when index errors are present compared to when they are not.
Additionally, when comparing scenarios with and without inter-group coupling, the effective GD STD remains nearly unchanged in the presence of index errors.
These findings indicate that the GD STD is primarily influenced by the contribution from the error penalty term. 
It is crucial for fabrication processes to maintain tight tolerances in order for GD compensation in MMF to be effective.

\subsection{Digital Signal Processing}
GD STD is a measure of the memory of the MMF channel, which in turn affects the DSP complexity of the MIMO equalizer. 
Additionally, since the MMF channel is dynamic, the equalization process must be adaptive \cite{arik_group_2016}. 
Submarine SMF channels have been observed to change on a time scale of seconds \cite{zhang_fast_2006}.
Fibers supporting more modes are expected to be more sensitive to environmental perturbations \cite{choutagunta_efficient_2019, choutagunta_modal_2019}.
A larger GD STD makes it more challenging to learn and adapt to a time-varying channel in the presence of noise.
The number of MIMO equalizer taps required at the receiver is given by \cite{kaminow_chapter_2013}
\begin{equation}
    N_{\mathrm{MD}} = \left\lceil \sigma_{\mathrm{GD}}~u_D(p)~r_{\textrm{os}} R_s \right\rceil,
\end{equation}
where $u_D(p)\approx 4.6$ for $D=12$ with probability $1-p=1-10^{-6}$, $r_{\textrm{os}}$ is the oversampling ratio and $R_s$ is the symbol rate. 
For example, for $\sigma_{\mathrm{GD}}=1~\textrm{ns}$, $r_{\textrm{os}}=17/16$, and $R_s = 50~\textrm{GBaud}$, we need $N_{\mathrm{MD}}=245$ taps. 
An adaptive $12\times 12$ MIMO frequency-domain equalization using a cyclic prefix (CP) of length  $N_{\mathrm{MD}}$ taps and block length of $N_{\textrm{FFT}} = 2^{12}$ achieves a CP efficiency $N_{\textrm{FFT}}/(N_{\textrm{FFT}}+N_{\mathrm{MD}}) = 94\%$. 

Adaptive algorithms such as recursive least-squares (RLS) can train the equalizers using $n_{\mathrm{tr}} = 50$ training blocks corresponding to an adaptation time of $t_\textrm{adapt} = (N_{\textrm{FFT}}+N_{\mathrm{MD}})/r_{\textrm{os}}R_s = 4.09~\mu\textrm{s}$.
The data-aided training stage can be followed by decision-directed tracking.
For a signal-to-noise ratio of $20~\textrm{dB}$ at the end of $100$ spans and a MDL STD of $0.3~\textrm{dB}$ per span, the RLS algorithm is found to track a dynamic channel with a generalized stokes rotation rate of more than $314~\textrm{krad/s}$ for QPSK signaling and up to $100~\textrm{krad/s}$ for 16-QAM signaling.
To determine the Stokes rotation rates and select an appropriate modulation and coding scheme, it is essential to experimentally measure the dynamics of the fabricated MMFs.
If the observed dynamics are faster than the estimates made here, then the FFT block length $N_{\textrm{FFT}}$ can be reduced and the channel estimates can be computed more frequently, although with a reduced CP efficiency.
Alternatively, pilot symbols can be periodically transmitted to enable intermittent data-aided tracking.
This method also leads to a reduced information rate.


\subsection{Extension to More Than Three Mode Groups}
\label{subsec:extension}

The analytical expressions presented in Section \ref{sec:Analytical} are applicable to all GD compensation systems characterized by strong random intra-group coupling and weak inter-group coupling in the fiber segments.

Fibers that support more than three mode groups can be designed using the parameterized refractive index profile \eqref{eq:ref_ind}.
However, additional features may be necessary to control the GDs and CDs of the mode groups. This is essential to creating a feature space sufficiently rich to enable numerical design optimization.

For the self-compensation scheme, the optimal mode permutation and the associated mode-group-averaged GD vector change as the number of mode groups increases. 
Combinatorially, there exist several possible ideal permutation schemes for any number of mode groups $N_g$. 
It is important to emphasize that realizing some of these ideal schemes may not be trivial, as each scheme requires a MMF with mode groups having specific GDs, as well as a mode permuter realizing a specific permutation pattern. We discuss two  examples here.

The first ideal permutation scheme consists of partitioning the mode groups into three groups, each with a different mode-group-averaged GD from the set $\{-\tau, 0, \tau\}$. 
The number of mode groups with a GD of 0 is $m = N_g \pmod 4$. For odd $m$, one of these mode groups has $(N_g+1)/2$ modes. For $m = 2,3$, there exists one pair of mode groups with a GD of $0$ comprising mode groups MG$_i$ and MG$_j$, where $i,j$ are the mode group indices and $i + j = N_g + 1$. 

The remaining number of mode groups is divisible by $4$, so we can form an even number of pairs of mode groups such that for every pair comprising MG$_i$, MG$_j$, $i + j = N_g + 1$. 
If the transmission fiber is designed so that half of these pairs have a GD of $\tau$ and the other half have a GD of $-\tau$, then permuting these two groups at the midpoint of each span effects perfect self-compensation (more generally, if $k_p>1$, we permute the two groups at the midpoint of each compensation period). 
For example, when $N_g=5$, the ideal GD vector for this scheme is $[\tau, -\tau, 0, -\tau, \tau]$. 
Midway through each span (or compensation period), MG$_1$ and MG$_5$ are permuted with MG$_2$ and MG$_4$, and MG$_3$ remains unchanged. 
For larger values of $N_g$, this scheme can be extended to allow different GD values for different pairs of mode groups, as long as for each pair of mode groups with a GD of $\tau$, there exists a corresponding mode group pair with a GD of $-\tau$.

An alternative ideal permutation scheme involves partitioning the mode groups into four groups, each with a different mode-group-averaged GD from the set $\{0, \tau_1, \tau_2, \tau_3\}$.
The number of mode groups with a GD of $0$ is $m = N_g \pmod 6$. For odd $m$, one of these mode groups has $(N_g+1)/2$ modes. For $m\geq2$, the other pairs of mode groups with a GD of $0$ are comprised of MG$_i$, MG$_j$, where $i + j = N_g + 1$. 

Since the remaining number of mode groups with nonzero GD is divisible by 6, we can form $M$ pairs of mode groups such that for every pair comprised of MG$_i$, MG$_j$, $i + j = N_g + 1$, where $M$ is divisible by 3. 
If the transmission fiber is then designed so that a third of these pairs have a GD of $\tau_1$, a third have a GD of $\tau_2$, and a third have a GD of $\tau_3$, cyclically permuting these three groups twice per span effects perfect self-compensation (more generally, if $k_p > 1$, we permute these three groups twice per compensation period). 
For example, in the case where $N_g=6$, the ideal GD vector for this scheme is $[\tau_1, \tau_2, \tau_3, \tau_3, \tau_2, \tau_1]$. 
We subdivide each span into three equal-length segments
(more generally, when $k_p > 1$, we subdivide each compensation period into three segments), and at the two boundaries between the three segments, we apply an ideal permutation that cycles power from MG$_1$ and MG$_6$ to MG$_2$ and MG$_5$, from MG$_2$ and MG$_5$ to MG$_3$ and MG$_4$, and from MG$_3$ and MG$_4$ to MG$_1$ and MG$_6$. Power can also be cycled amongst these three pairs of mode groups in the reverse direction. 
For larger values of $N_g$, multiple such 3-cycles can exist, each with potentially different values of $\tau_1, \tau_2, \tau_3$.

Although these ideal permutation schemes have been shown to exist for any $N_g$, their implementation becomes increasingly difficult as $N_g$ increases. 
Designing a mode permuter for large $N_g$ requires careful consideration, especially when opting for cascaded LPFG-based devices.
The sheer number of necessary power transfers amongst all modes greatly increases the number of LPFGs the device requires. 
In the schemes described here, potentially one mode would need to be transferred to a mode group four levels away, and to prevent coupling into unguided modes, this transfer should take place over several steps, each requiring an LPFG \cite{krutko_low-loss_2025}. 
There is likely a limit to how many different LPFGs can be inscribed in the same fiber without introducing significant undesired coupling, so multiple mode permutation fibers may be necessary \cite{krutko_low-loss_2025}. 
These are topics for future research.

\section{Conclusion}
\label{sec:Conclusion}
In this paper, we examined GD compensation for MMF links using two approaches: conventional compensation, which alternates fiber types with opposite GD orderings, and self-compensation, which utilizes a single fiber type with periodically inserted mode permuters. 
We provided analytical expressions for the GD STD in these systems, taking into account factors such as random inter-group coupling, mode scrambling, and refractive index errors. 
We proposed fiber designs based on graded-index MMFs with specialized index profiles to control the MDCD in order to achieve effective compensation over the entire C-band.
We also addressed design strategies for both systems and fibers to address the challenges of random inter-group coupling and refractive index errors. 
These strategies include increasing the number of compensation periods per span, using a regularized cost function in fiber design, and fiber sorting.
The design examples we presented achieved an effective GD STD of $18.1~\textrm{ps}/\sqrt{\textrm{km}}$ with conventional compensation and $13.3~\textrm{ps}/\sqrt{\textrm{km}}$ with self-compensation for $\linter = 500~\textrm{km}$ and $\sigDeln=10^{-5}$.  
We discussed the implications of GD compensation for DSP complexity and performance. Finally, we discussed  methods for extending GD compensation  to fibers that support more than three mode groups.

\appendices
\section{Impact of Refractive Index Errors on Modal Group Delays}
\label{app:error_derivation}

In this appendix, we derive a semi-analytic expression for the penalty in GD STD caused by errors in the fiber refractive index profile.
Consider a fiber with a refractive index profile $n(r)$ and
refractive index error $\Delta n(r)$ as a function of radial position $r$.
From a first-order perturbation theory analysis, the change in the propagation constant of the $i$th mode is
\begin{equation}
    \Delta \beta[i] = \frac{1}{2\beta[i]}\frac{\iint 2 k_0^2 n \Delta n |\psi_i|^2 rdrd\theta}{\iint |\psi_i|^2 rdrd\theta},
\end{equation}
where $\psi_i$ is the electric field amplitude of the $i$th mode. The error in its GD is
\begin{equation}
    \Delta \beta_1[i] = \frac{d}{d\omega}\Delta\beta[i].
\end{equation}

Numerically, the integration can be replaced by a summation and the derivative can be replaced by a finite difference as follows: 
\begin{equation}
    \Delta \beta_1[i] = \frac{1}{2}\sum_j \frac{1}{\Delta \omega}\left(\frac{a_j^+[i]}{\beta^+[i]} - \frac{a_j^-[i]}{\beta^-[i]}\right)\Delta n[j],
\end{equation}
where $j=1,\dots,N_r$ is a radial position index, $\Delta \omega$ is a frequency difference, superscripts $+$ and $-$ indicate the values at $+\Delta\omega/2$ and $-\Delta\omega/2$, respectively, and
\begin{equation}
    a_j[i] = 2 k_0^2 n[j] r[j]\frac{|\psi_i[j]|^2}{\sum_l |\psi_i[l]|^2 r[l] }.
\end{equation}
From the above expressions, the errors in the mode-group-averaged GDs can be obtained:
\begin{equation}
    \Delta\barbeta_1 = \mathbf{\bar{\textbf{A}}} \Delta\mathbf{n},
\end{equation}
where 
$\Delta\mathbf{n}$ is the $N_r$-dimensional error vector and $\mathbf{\bar{\textbf{A}}}$ is an $N_g\times N_r$ matrix describing the mode-group-averaged GD error coefficients and is given by 
\begin{equation*}
    \mathbf{\bar{\textbf{A}}}[l,j] = \frac{1}{2d_l \Delta \omega}\sum_{i\in\mathcal{M}_l}{\left(\frac{a_j^+[i]}{\beta^+[i]} - \frac{a_j^-[i]}{\beta^-[i]}\right)}.
\end{equation*}

Assuming $\{\Delta n[j]\}$ are independent and identically distributed Gaussian random variables with STD $\sigDeln$ results in 
$\Delta\barbeta_1$ being a zero-mean Gaussian random vector whose second-order statistics are related to those of $\Delta \mathbf{n}$ by
\begin{equation}
    \Sigma = \E{\Delta\barbeta_1^T \Delta\barbeta_1} = \sigDeln^2 \bar{\textbf{A}}\bar{\textbf{A}}^H .
\end{equation}


\section*{Acknowledgment}
This project was supported by Ciena Corporation, Stanford Shoucheng Zhang Graduate Fellowship, and Stanford Graduate Fellowship.
Much of the computing for this project was performed on the Sherlock cluster at Stanford University. 
We thank the Stanford Research Computing Center for providing this cluster and technical support.
The authors are grateful for helpful discussions with Dr. Ethan Liang.

\bibliographystyle{IEEEtran}
\bibliography{references}

\begin{IEEEbiographynophoto}{Anirudh Vijay}
received the B.Tech. and M.Tech. degrees in Electrical Engineering from the Indian Institute of Technology Madras, Chennai, Tamil Nadu, India, in 2019.
He is working towards the Ph.D. degree in Electrical Engineering from Stanford University, Stanford, CA, USA. 
His current research interests include optical communications, mode-division multiplexing, and data-center applications.
\end{IEEEbiographynophoto}

\begin{IEEEbiographynophoto}{Nika Zahedi}
is currently working towards the B.S. and M.S. degrees in electrical engineering at Stanford University. Her current research interests include mode-division multiplexing, signal processing, and optimization techniques.
\end{IEEEbiographynophoto}

\begin{IEEEbiographynophoto}{Oleksiy Krutko}
received the B.S. degree in electrical engineering from the University of Texas at Austin, Austin, TX, USA, in 2020. He is currently working toward the Ph.D. degree from Stanford University, Stanford, CA, USA. His research interests include optical fiber communications and photonic devices.
\end{IEEEbiographynophoto}

\begin{IEEEbiographynophoto}{Rebecca Refaee}
received a B.S. degree in Mathematics and an M.S. degree in Electrical Engineering from Stanford University, Stanford, CA, USA in 2024. 
She is currently working towards a Ph.D. degree in Electrical Engineering at Stanford University. Her current research interests include optical communications and mode-division multiplexing.
\end{IEEEbiographynophoto}

\begin{IEEEbiographynophoto}{Joseph M. Kahn}
(F’00) received A.B., M.A. and Ph.D. degrees in Physics from the University of California, Berkeley in 1981, 1983 and 1986. In 1987-1990, Kahn was at AT\&T Bell Laboratories. In 1989, he demonstrated the first successful synchronous (i.e., coherent) detection using semiconductor lasers, achieving record receiver sensitivity. In 1990-2003, Kahn was on the Electrical Engineering and Computer Sciences faculty at Berkeley. He demonstrated coherent detection of QPSK in 1992. In 1999, D. S. Shiu and Kahn published the first work on probabilistic shaping for optical communications. In the 1990s and early 2000s, Kahn and collaborators performed seminal work on indoor and outdoor free-space optical communications and multi-input multi-output wireless communications. In 2000, Kahn and K. P. Ho founded StrataLight Communications, whose 40 Gb/s-per-wavelength long-haul fiber transmission systems were deployed widely by AT\&T, Deutsche Telekom, and other carriers. In 2002, Ho and Kahn applied to patent the first electronic compensation of fiber Kerr nonlinearity. StrataLight was acquired by Opnext in 2009. In 2003, Kahn became a Professor of Electrical Engineering in the E. L. Ginzton Laboratory at Stanford University. Kahn and collaborators have extensively studied rate-adaptive coding and modulation, as well as digital signal processing for mitigating linear and nonlinear impairments in coherent systems. In 2008, E. Ip and Kahn (and G. Li independently) invented simplified digital backpropagation for compensating fiber Kerr nonlinearity and dispersion. Since 2004, Kahn and collaborators have studied propagation, modal statistics, spatial multiplexing and imaging in multimode fibers, elucidating principal modes and demonstrating transmission beyond the traditional bandwidth-distance limit in 2005, deriving the statistics of coupled modal group delays and gains in 2011, and deriving resolution limits for imaging in 2013. Kahn’s current research addresses optical frequency comb generators, coherent data center links, rate-adaptive access networks, fiber Kerr nonlinearity mitigation, ultra-long-haul submarine links, and optimal free-space transmission through atmospheric turbulence. Kahn received the National Science Foundation Presidential Young Investigator Award in 1991. In 2000, he became a Fellow of the IEEE.
\end{IEEEbiographynophoto}

\end{document}